\documentclass[fleqn,usenatbib]{mnras}

\usepackage{orcidlink}
\usepackage{longtable}
\usepackage{supertabular} 
\usepackage{lscape}
\usepackage{lipsum}
\usepackage{newtxtext,newtxmath}
\usepackage{multirow}
\usepackage[T1]{fontenc}

\DeclareRobustCommand{\VAN}[3]{#2}
\let\VANthebibliography\thebibliography
\def\thebibliography{\DeclareRobustCommand{\VAN}[3]{##3}\VANthebibliography}

\usepackage{graphicx}	% Including figure files
\usepackage{amsmath}	% Advanced maths commands
\title[radio-gamma-ray properties]{Radio–Gamma-Ray Properties and High-Energy Implications for Fermi Blazars}

\author[Ye et al.]{
Xu-Hong Ye\orcidlink{0000-0003-2551-0202},$^{1}$ Wen-Xin Yang\orcidlink{0000-0001-8244-1229},$^{1,2,3,4}$
Guo-Hai Chen\orcidlink{0009-0006-2844-8677},$^{1,2,3,4}$ Zhi-Yuan Pei\orcidlink{0000-0002-4970-3108},$^{1,3,4}$ Yong-Yun Chen\orcidlink{0000-0001- 5895-0189},$^{5}$
\newauthor 
Yi Liu\orcidlink{0000-0003-3863-9777},$^{1,3,4}$
Denis Bastieri\orcidlink{0000-0002-6954-8862},$^{2,6,1}$ and
Jun-Hui Fan\orcidlink{0000-0002-5929-0968}$^{1,3,4}$ \thanks{E-mail: fjh@gzhu.edu.cn}
\\
$^{1}$Center for Astrophysics, Guangzhou University, Guangzhou 510006, People's Republic of China\\
$^{2}$Dipartimento di Fisica e Astronomia "G. Galilei", Università di Padova, Via F. Marzolo, 8, I-35131 Padova, Italy\\
$^{3}$Astronomy Science and Technology Research Laboratory of Department of Education of Guangdong Province, Guangzhou 510006, People's Republic of China\\
$^{4}$Greater Bay Brand Center of the National Astronomical Data Center, Guangzhou 510006, People's Republic of China\\
$^{5}$College of Physics and Electronic Engineering, Qujing Normal University, Qujing 655011, People's Republic of China \\
$^{6}$ Istituto Nazionale di Fisica Nucleare, Sezione di Padova, I-35131 Padova, Italy\\
}

\date{Accepted XXX; Received XXX; in original form XXX}

\pubyear{2026}

\begin{document}
\label{firstpage}
\pagerange{\pageref{firstpage}--\pageref{lastpage}}
\maketitle

\begin{abstract}
Radio and $\gamma$-ray emissions in blazars, a subclass of active galactic nuclei (AGNs), provide important insight into their high-energy radiation processes. We studied the relation between radio and $\gamma$-ray emissions using a large sample of 1687 \textit{Fermi} blazars, based on the Radio Fundamental Catalogue and the latest Third Data Release of the Fourth \textit{Fermi} AGN Catalogue. A clear correlation between radio and $\gamma$-ray fluxes for both BL Lacertae objects (BL Lacs) and flat-spectrum radio quasars (FSRQs) suggests a synchrotron self-Compton (SSC) contribution to both subclasses. The ratio of $\gamma$-ray and radio emissions, $\gamma$-ray loudness ($G_{\rm r}$), is further examined with the $\gamma$-ray photon index ($\Gamma_\gamma$) and the synchrotron peak frequency ($\nu_{\rm{peak}}$). An anti-correlation between $G_{\rm r}$ and $\Gamma_\gamma$ is explained by the shift of the spectral energy distribution rather than the Compton cooling effect.  We found that $G_{\rm r}$ shows a positive dependence on $\nu_{\rm{peak}}$ for low-synchrotron-peaked BL Lacs (LBLs) and FSRQs, in line with the SSC-contributed scenario, although additional external Compton contributions may account for the substantial scatter observed in LBLs and FSRQs. In contrast, high-synchrotron-peaked BL Lacs (HBLs) reach the plateau of $G_{\rm r}$ between $\log (\nu_{\rm peak}/{\rm Hz}) \simeq15.5-16$, possibly indicating the transition from the Thomson to the Klein--Nishina (KN) regime. Interpreting this feature within a one-zone SSC framework could constrain the magnetic field strength of $-4.14 < \log (B/{\rm G}) < -1.69$ for those HBLs affected by the KN suppression.

\end{abstract}


%%%%%%%%%%%%%%%%%%%%%%%%%%%%%%%%%%%%%%%%%%%%%%%%%%

%%%%%%%%%%%%%%%%% BODY OF PAPER %%%%%%%%%%%%%%%%%%

\section{Introduction}\label{intro}
Blazars are a subclass of radio-loud active galactic nuclei (RLAGNs) with their relativistic jets pointing close to the line of sight, which can be divided into two types based on the optical equivalent width (EW):  BL Lacertae objects (BL Lacs) with weak or no emission lines (EW $<5 \mathring{A}$), or flat-spectrum radio quasars (FSRQs) with strong emission lines (EW $>5 \mathring{A}$) \citep{stickel.1991.apj.374,urry.1995.pasp.107,falomo14}. 

The observational properties and the characteristic double-peaked spectral energy distributions (SEDs) of blazars are dominated by the relativistic jet \citep{ghisellini2010mnras,blandford19,elisa2022,ulgiati25}. The low-energy peak, ranging from radio to X-rays, arises from synchrotron radiation, while the high-energy peak is produced by inverse Compton (IC) scattering of low-energy seed photons by relativistic electrons. The origin of these seed photons strongly affects the observed $\gamma$-ray emissions \citep{finke2008apj,ghisellini2010mnras,cerruti2020galaxies}. When seed photons are internal to the jet, e.g., in BL Lacs, the high-energy emission follows the synchrotron self-Compton (SSC) emission. In contrast, external seed photons from the broad-line region, dusty torus, accretion disc, or cosmic microwave background are scattered by relativistic electrons, e.g., in FSRQs, leading to the external Compton (EC) emission \citep{dermer93apj,sikora94apj,blazejowski00apj,finke2008apj,abdo_sci_2010_cena}. Therefore, BL Lacs and FSRQs are expected to exhibit different $\gamma$-ray beaming patterns in the high-energy scenarios \citep{dermer1995apjl,sikora09apj,fan2013PASJ,yang22raa,ye26}.   

The relativistic jet in blazars not only produces the double-peak phenomenon, but also shapes a sequence in SEDs \citep[e.g.,][]{fossati98mnras,elisa2022}: an anti-correlation between the synchrotron peak frequency ($\nu_{\rm{peak}}$) and the radio (or $\gamma$-ray) emissions. The trend is extensively discussed, which could be explained by the Compton cooling effect \citep{fossati98mnras,ghisellini98,elisa2022,ye25}, selection effects of the sample \citep{nie2008A&A,chen11apj,giommi2012mnras,wan24mn}, mass accretion rate \citep{boula19mn,boula26} or beaming effects \citep{fan17apjl,yangwx2022apj,oyzh2023ApJ}. 
The location of the $\nu_{\rm{peak}}$ also classifies RLAGNs into low-synchrotron-peaked (LSP, $\log (\nu_{\rm peak}/\rm{Hz})< 14$), intermediate-synchrotron-peaked (ISP, $14 \le \log (\nu_{\rm peak}/\rm{Hz})< 15$), high-synchrotron-peaked (HSP, $15\le\log (\nu_{\rm peak}/\rm{Hz})\le 17$), and extreme high-synchrotron-peaked (EHSP, $\log (\nu_{\rm peak}/\rm{Hz})\ge 17$) sources \citep{abdo10sed,fan16,maria25aa,sibani26aa,abe26apj}. In general, FSRQs are homogeneous and mostly correspond to LSPs, whereas BL Lacs are heterogeneous, including EHBLs, HBLs, IBLs, and LBLs (extreme high-, high-, intermediate-, and low-synchrotron-peaked BL Lacs) \citep{costamante01aa,nie06,foffano19mn,ajello.2022.apjs.263,yjh22apjs,lian26apj}.

Since the launch of the \textit{Fermi} Large Area Telescope (hereafter \textit{Fermi}), blazars have been detected to dominate the extragalactic $\gamma$-ray sky \citep{abdo10apj429,abdollahi.apjs.2022.260}. The latest Third Data Release of the Fourth \textit{Fermi} AGN Catalogue (4LAC-DR3) contains 3814 $\gamma$-ray detected AGNs, among which 3743 AGNs ($98.14\%$) are classified as blazars, including 792 FSRQs, 1458 BL Lacs, and 1493 blazar candidates of uncertain type (BCUs) \citep{ajello.2022.apjs.263,abdollahi.apjs.2022.260}. Their $\gamma$-ray emissions are closely linked to their radio properties, exhibiting higher radio brightness temperature,  stronger core-dominance phenomena, significant apparent superluminal motions, strongly
polarized jets and larger Doppler factors  \citep{moe96aj,lister05,kovalev09apjl,lister09ApJL,linford12apj,wu14,pzy2020scpma,homan21ApJ,xiao2022mnras}.  The strong correlation between radio and $\gamma$-ray emissions in blazars has been studied, showing that
the high-energy output is closely linked to the relativistic jets responsible for compact radio emissions \citep{fan98aa,fossati98mnras,kovalev09apjl,ojha10A&A,lister11apj,fanxuliang2012raa,linford12apj,bock16aa,cui26}. Very Long Baseline Interferometry (VLBI) provides the highest angular resolution, which can help further resolve the inner jets of blazars and measure their jet kinematics \citep{deller14,bock16aa,petrov25apjs}. The connection between VLBI and \textit{Fermi} motivates one to study the jet properties of $\gamma$-ray blazars, and understand their high-energy radiation mechanisms. In this paper, we utilised a large sample combining the RFC (Radio Fundamental Catalogue) VLBI catalogue\footnote{https://astrogeo.org/sol/rfc/rfc\_2024d \citep{petrov25apjs}.} and the \textit{Fermi} 4LAC-DR3 AGN catalogue \citep{ajello.2022.apjs.263} to discuss the radio and $\gamma$-ray properties, as well as the high-energy mechanisms.

\section{Sample}

To investigate the parsec-scale radio properties of $\gamma$-ray blazars, we cross-matched the 4LAC-DR3 sample with the RFC catalogue compiled by \citet{petrov25apjs}, who present flux density measurements for 21942 compact radio sources based on the reanalysis of 1088 VLBI experiments conducted between 1994 and 2024. The observations cover the precise positions and the flux
densities at 2.2–2.4 GHz (S band), 4.1–5 GHz (C band), 7.3–8.8 GHz (X band), 15.2–15.5 GHz (U band), and 22–24.2 GHz (K band). There are three different baseline lengths for the VLBI detection: median flux density at the baseline projection lengths shorter than 1000 km, between 1000 km and 5000 km, and longer than 5000 km.

The cross-matching process between the \textit{Fermi} 4FGL-DR3 and RFC catalogues yields 2578 $\gamma$-ray blazars, including 1057 BL Lacs, 735 FSRQs, and 786 BCUs. The VLBI flux densities are obtained at C band or X band with baseline projections longer than 5000 km, which corresponds to the unresolved flux and refers to the compact core emission on the parsec scale for blazars. This results in a blazar sample of confirmed 976 BL Lacs and 711 FSRQs with the available VLBI radio core emissions.
The compact VLBI radio spectrum for RLAGNs in the GHz band is flat due to synchrotron self-absorption (e.g., \citealt{blandford79apj,lobanov98aa,abdo10sed,grandi25}); therefore, these measurements between C band (4.1 - 5 GHz) and X band (7.3 - 8.8 GHz) provide a reliable and consistent compact radio core emission for blazars.  

Overall, the sample includes 976 BL Lacs and 711 FSRQs with both the VLBI and $\gamma$-ray parameters.  
Although the sample is not statistically complete from both the VLBI and \textit{Fermi} surveys, compared with earlier radio and $\gamma$-ray studies based on a few hundred sources (e.g., \citealt{lister11apj, linford12apj, bock16aa}), its large sample size and broad coverage in radio flux, $\gamma$-ray flux, $\nu_{\rm{peak}}$, and spectral properties make it representative of the bright $\gamma$-ray blazar population. Therefore, selection biases are not expected to strongly affect our results.

\begin{table*}
\centering
\caption{Radio and $\gamma$-ray properties for 1687 \textit{Fermi} blazars. }
\label{tab:multiwave_properties}
\begin{tabular}{cccccccc}
\hline
Source Name & Class & $\log \nu_{\rm peak}$& $\Gamma_{\gamma}$& $\log f_\gamma$ & VLBI Counterpart & $\log S_{\rm{VLBI}}$ & $\log G{\rm{r}}$ \\
(1)&(2)&(3)&(4)&(5)&(6)&(7)&(8)\\
\hline
4FGL J0001.2-0747	&	BL Lac	&	13.96 	&	2.12 	&	-11.08 	&	RFC J0001-0746	&		-1.29 		&	3.51 	\\
4FGL J0001.5+2113	&	FSRQ	&	14.20 	&	2.65 	&	-10.59 	&	RFC J0001+2113	&		-1.18 		&	3.89 	\\
4FGL J0003.2+2207	&	BL Lac	&	13.92 	&	2.12 	&	-12.19 	&	RFC J0003+2204	&		-2.10 		&	3.20 	\\
4FGL J0003.9-1149	&	BL Lac	&	12.39 	&	2.18 	&	-11.63 	&	RFC J0004-1148	&	    -0.69 		&	2.37 	\\
4FGL J0004.0+0840	&	BL Lac	&	14.48 	&	2.00 	&	-11.94 	&	RFC J0003+0841	&		-1.89 		&	3.25 	\\
4FGL J0004.3+4614	&	FSRQ	&	12.41 	&	2.59 	&	-11.29 	&	RFC J0004+4615	&		-0.81 		&	2.83 	\\
4FGL J0004.4-4737	&	FSRQ	&	13.12 	&	2.44 	&	-11.27 	&	RFC J0004-4736	&		-0.68 		&	2.71 	\\
4FGL J0005.9+3824	&	FSRQ	&	13.12 	&	2.62 	&	-11.19 	&	RFC J0005+3820	&		-0.92 		&	3.03 	\\
4FGL J0006.3-0620	&	BL Lac	&	12.92 	&	2.15 	&	-11.85 	&	RFC J0006-0623	&		0.37 		&	1.08 	\\
4FGL J0008.0+4711	&	BL Lac	&	13.52 	&	2.05 	&	-10.73 	&	RFC J0007+4712	&		-1.70 		&	4.27 	\\
4FGL J0008.4-2339	&	BL Lac	&		&	1.72 	&	-11.67 	&	RFC J0008-233A	&		-1.82 		&	3.46 	\\
4FGL J0009.1+0628	&	BL Lac	&	13.39 	&	2.15 	&	-11.23 	&	RFC J0009+0628	&		-1.03 		&	3.10 	\\
4FGL J0009.3+5030	&	BL Lac	&	15.10 	&	1.99 	&	-10.64 	&	RFC J0009+5030	&		-2.05 		&	4.71 	\\
4FGL J0010.6+2043	&	FSRQ	&	12.56 	&	2.39 	&	-11.71 	&	RFC J0010+2047	&		-1.28 		&	2.87 	\\
4FGL J0010.6-3025	&	FSRQ	&	12.99 	&	2.43 	&	-11.27 	&	RFC J0010-3027	&		-0.73 		&	2.76 	\\
4FGL J0011.4+0057	&	FSRQ	&	12.84 	&	2.35 	&	-11.13 	&	RFC J0011+0057	&		-0.91 		&	3.08 	\\
4FGL J0013.0+3355	&	FSRQ	&	12.50 	&	2.26 	&	-11.75 	&	RFC J0012+3353	&		-0.89 		&	2.44 	\\
4FGL J0013.1-3955	&	BL Lac	&	12.86 	&	2.05 	&	-11.38 	&	RFC J0012-3954	&		-0.31 		&	2.22 	\\
$\ldots$&$\ldots$&$\ldots$&$\ldots$&$\ldots$&$\ldots$&$\ldots$&$\ldots$\\
\hline
\end{tabular}

\small
Note: Columns list (1) source name, (2) class, (3) logarithmic synchrotron peak frequency ($\nu_{\rm peak}$), 
(4) $\gamma$-ray photon index ($\Gamma_{\gamma}$), (5) logarithmic gamma-ray energy flux, $\log f_\gamma$ (in units of erg/cm$^2$/s), 
(6) VLBI counterpart, 
(7) logarithmic VLBI flux density, $\log S_{\rm{VLBI}}$ (in units of Jy), and (8) logarithmic ratio $G_{\rm{r}}=f_{\gamma}/{\nu S_{\rm{VLBI}}}$. 

A portion of this table is shown here for guidance. The full table is available in Zenodo, at https://doi.org/10.5281/zenodo.21695209 [doi].
\end{table*}

\subsection{Radio and $\gamma$-ray properties}
%In our sample, the radio flux density is derived from VLBI observations at 4.4 or 8 GHz with projected baselines longer than 5000 km, isolating the compact parsec-scale jet emission \citep{petrov25apjs}. 
The VLBI observations cover a wide range in flux density for the 1687 blazars, from $S_{\rm{VLBI}}=0.004$ Jy to $S_{\rm{VLBI}}=9$ Jy.
 The $\gamma$-ray average energy flux is taken from the 4FGL-DR3, which is integrated over the 0.1--100 GeV band \citep{ajello.2022.apjs.263}, ranging from $\log f_{\gamma}=-9.07$ erg/cm$^2$/s to $\log f_{\gamma}=-12.43$ erg/cm$^2$/s for  1687 blazars. 
 
\begin{figure}
    \centering
    \includegraphics[width=0.95\linewidth]{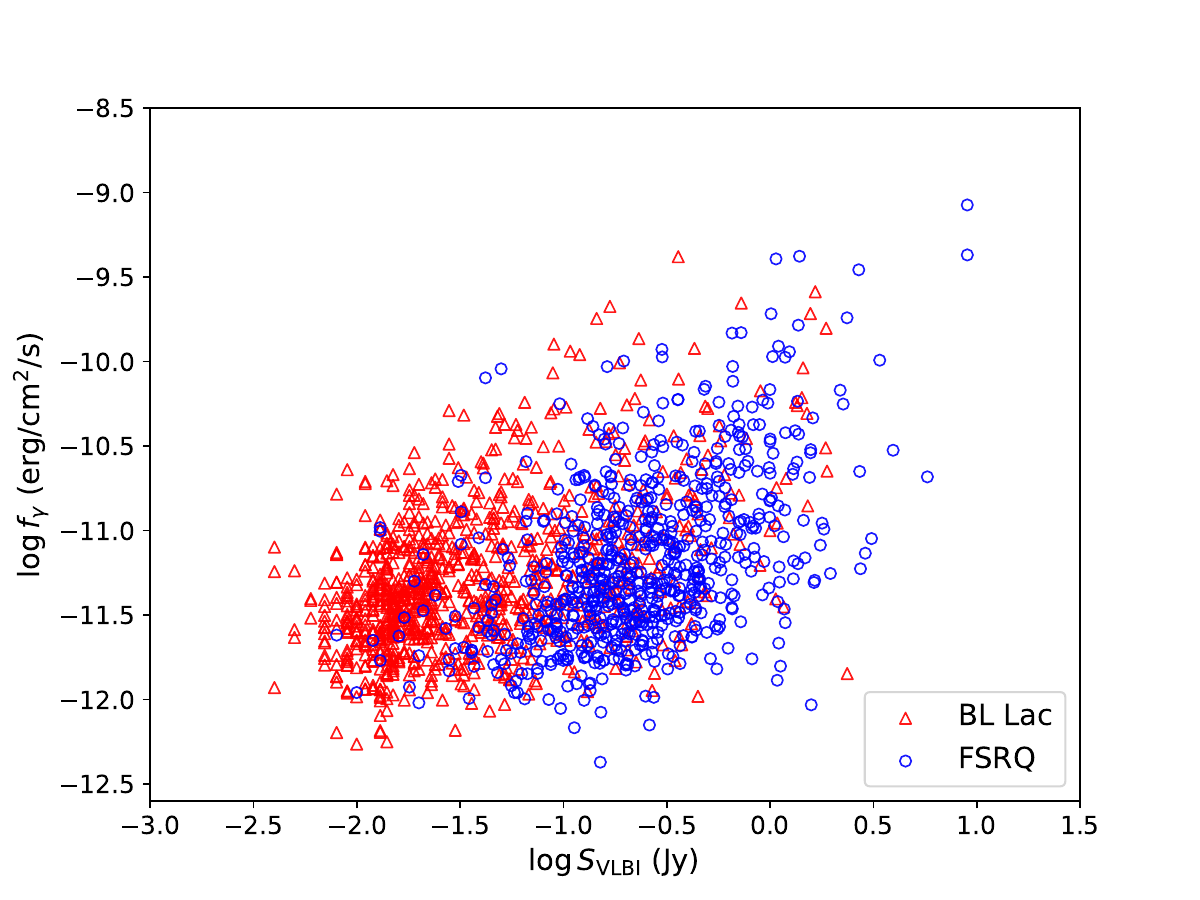}
    \caption{$\gamma$-ray energy flux ($ f_{\gamma}$) versus VLBI radio flux density ($S_{\rm{VLBI}}$) for \textit{Fermi} blazars.  Red triangles represent BL Lacs and blue circles represent FSRQs.}
    \label{fig:radio_gamma_flux}
\end{figure}

The average radio and $\gamma$-ray emissions for FSRQs ($ S_{\rm{VLBI}}=0.41\pm 0.03$ Jy; $\log f_{\gamma}=-11.18\pm 0.02$ erg/cm$^2$/s) are higher than those of BL Lacs ($S_{\rm{VLBI}}=0.10\pm 0.01$ Jy; $\log f_{\gamma}=-11.30\pm 0.01$ erg/cm$^2$/s), with the clear differences determined by the Kolmogorov–Smirnov (K-S) tests of the probability of $p<0.05$.  Figure~\ref{fig:radio_gamma_flux} shows the relation between the VLBI radio flux and the $\gamma$-ray energy flux. This clear positive correlation is confirmed by the Kendall $\tau$ test correlation coefficients of $r = 0.48$ and $p <10^{-4}$ for BL Lacs, and $r = 0.47 $ and $p <10^{-4} $ for FSRQs.
 
 \citet{lister09ApJL} discussed that the fainter VLBA\footnote{Very Long Baseline Array.} radio luminosities of the BL Lacs require a higher intrinsic (unbeamed) $\gamma$-ray to radio luminosity ratio than FSRQs, in order to account for their higher \textit{Fermi}-detection rate (48$\%$ detection rate for BL Lacs, while 20\% for FSRQs in the MOJAVE\footnote{Monitoring Of Jets in Active galactic nuclei with VLBA Experiments.} sample). With the large sample, the 976 BL Lacs show a lower average VLBI radio emission than the 711 FSRQs, in line with the idea of \citet{lister09ApJL} that a higher intrinsic (unbeamed) $\gamma$-ray to radio luminosity ratio might be involved for BL Lacs' detection, which could be further investigated by the unified scenario between blazars and radio galaxies in the future.

\subsection{$\gamma$-ray loudness and photon index}

The compact radio core emission traces the synchrotron emission from the relativistic jet, while the $\gamma$-ray emission is generally attributed to IC processes \citep{ghisellini11mnras,cerruti2020galaxies}. Differences in dominant radiative mechanisms (e.g., SSC versus EC) are expected to produce distinct $\gamma$-ray loudness distributions. The $\gamma$-ray loudness is defined as in \citet{lister11apj,linford12apj},
\begin{equation}
G_{\rm r} = \frac{\nu L_{\gamma}}{\nu L_{\rm VLBI}} 
           = \frac{f_{\gamma}}{\nu S_{\rm{VLBI}}},
\end{equation}
where $L_{\gamma}$ and $L_{\rm{VLBI}}$ are the $\gamma$-ray and VLBI luminosities, respectively. Here  $f_{\gamma}$ is the $\gamma$-ray energy flux and $S_{\rm{VLBI}}$ is the monochromatic radio flux density. 
Because the radio and $\gamma$-ray luminosities scale similarly with luminosity distance, $G_{\rm r}$ is insensitive to redshift effects. This makes it particularly useful for BL Lacs with uncertain or missing redshift measurements, which are often excluded from samples requiring redshift information (e.g., \citealt{fanxuliang18apj,maria25aa}). Cross-matching the redshift from the 4LAC-DR3 catalogue shows that 357 out of 976 BL Lacs would be excluded due to the lack of redshift measurements. As a result, our sample includes a larger fraction of BL Lacs without measured redshifts than previous studies. This allows for a more complete comparison between BL Lacs and FSRQs in both radio and $\gamma$-ray properties.

All sources in our sample are $\gamma$-ray loud, with $\log G_{\rm r}$ spanning $\log G_{\rm r}=1.07$ to $\log G_{\rm r}=4.70$ for 1687 blazars. \citet{linford12apj} showed a higher range of $\log G_{\rm r}=2.74$--$5.27$ for the 1FGL (\textit{Fermi} First Source Catalogue) sample. This difference likely arises from the different temporal coverage of the $\gamma$-ray data. The 1FGL catalogue is biased toward bright sources, whereas the 12-year averaged data would smooth out strong variability and reduce the contribution from flaring states, leading to lower mean $\gamma$-ray fluxes and smaller $G_{\rm r}$ values.

\citet{lister11apj} reported a clear separation in the distribution of $G_{\rm r}$ between BL Lacs and FSRQs, whereas \citet{linford12apj} found no difference between the two populations. \citet{cui26}  adopted the latest Nanshan 26-m radio telescope to observe \textit{Fermi} blazars and obtained a slightly higher $\gamma$-ray loudness in BL Lacs than FSRQs. 
In this sample, the FSRQs are less $\gamma$-ray loud, and BL Lacs show a slightly higher average $\gamma$-ray loudness. The average $\gamma$-ray loudness for BL Lacs ($\log G_{\rm {r}}=3.43\pm0.02$) is larger than that of FSRQs ($\log G_{\rm {r}}=2.75\pm0.02$),  with a significant difference of the K-S probability $p<0.05$.  

\begin{figure}
    \centering
    \includegraphics[width=0.95\linewidth]{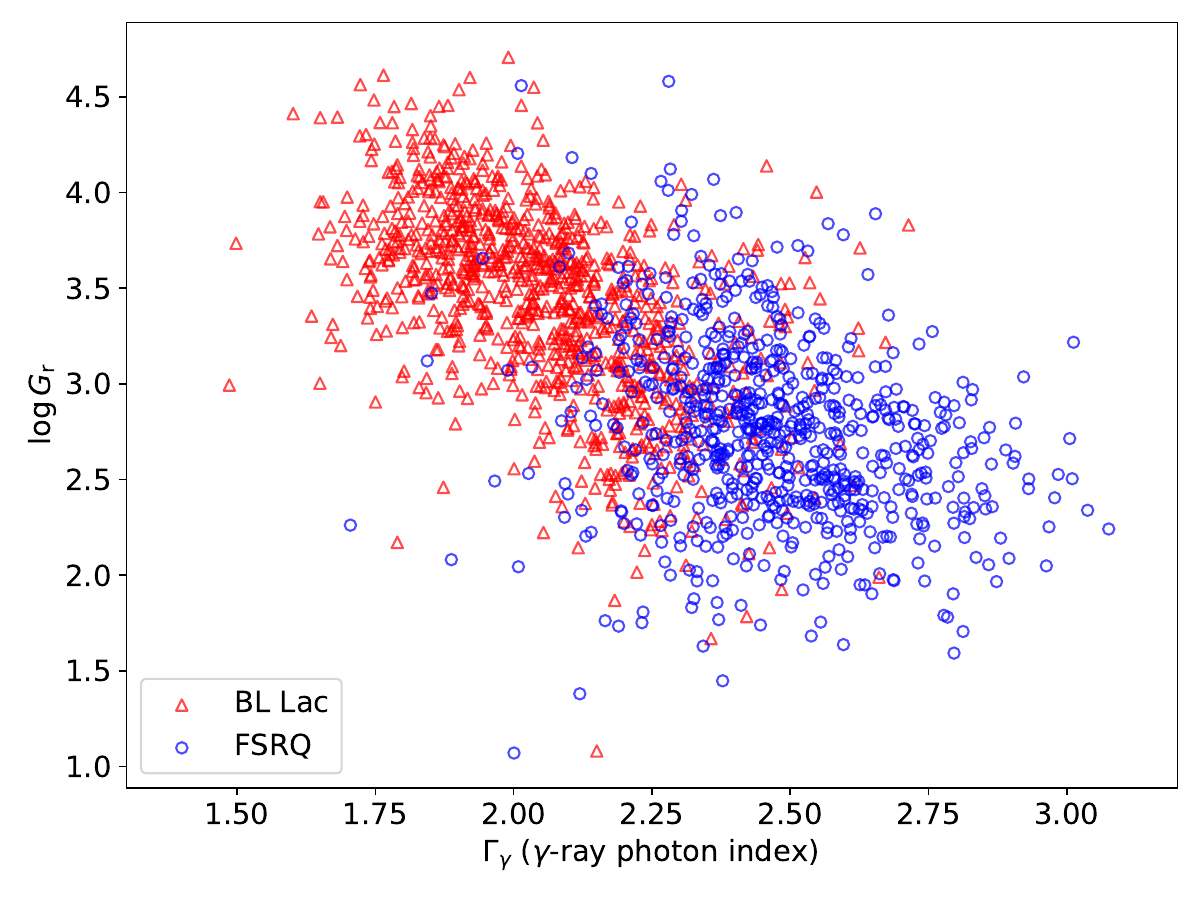}
    \caption{$\gamma$-ray loudness $G_{\rm r}$ versus $\gamma$-ray photon index $\Gamma_\gamma$ for \textit{Fermi} blazars.
    Symbols are the same as in Figure~\ref{fig:radio_gamma_flux}. 
    BL Lacs occupy regions of higher $G_{\rm r}$ and harder spectra compared to FSRQs.}
    \label{fig:gr_gamma_index}
\end{figure}
\begin{figure} 
    \centering
    \includegraphics[width=0.95\linewidth]{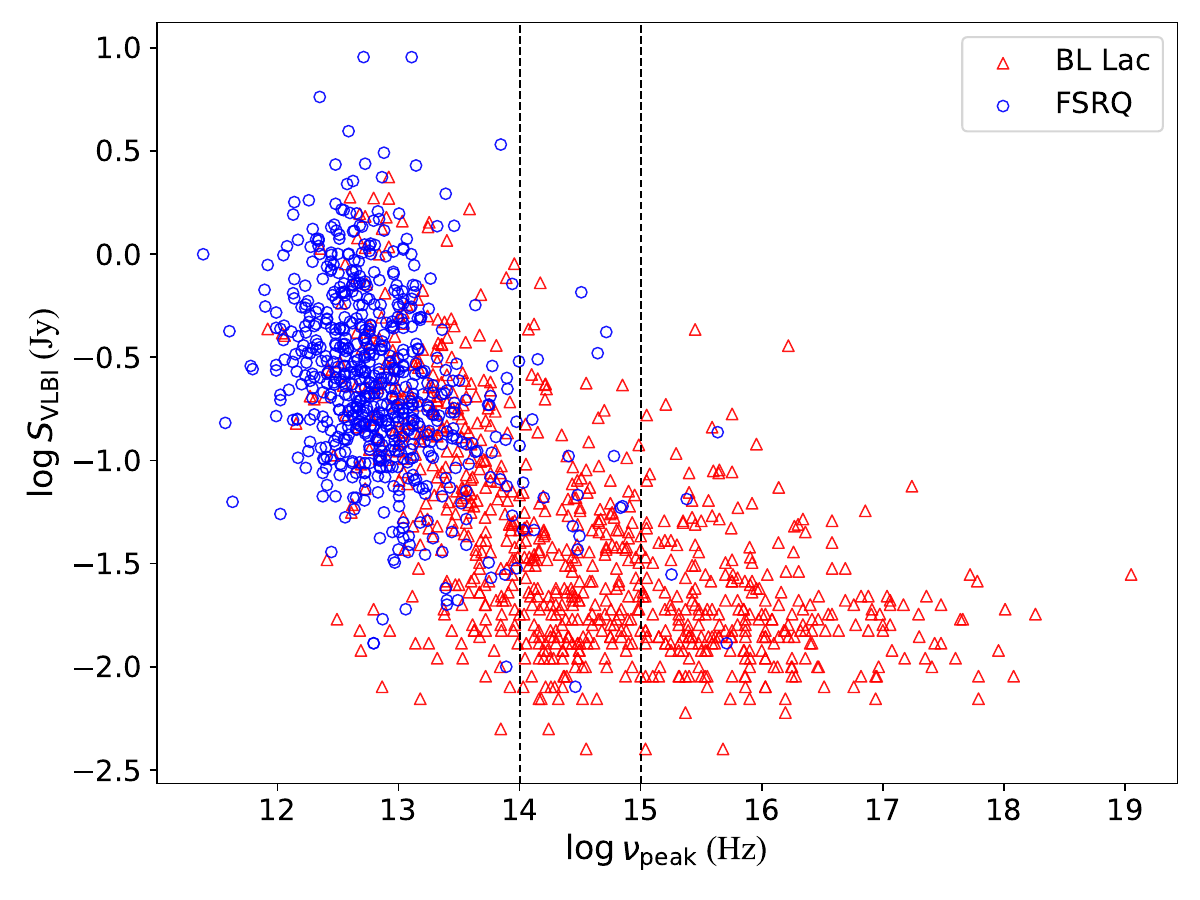}
    \caption{VLBI flux density ($S_{\rm{VLBI}}$) versus synchrotron peak frequency $\nu_{\rm{peak}}$ for \textit{Fermi} blazars.  
    The dashed vertical lines indicate the boundaries between LSP, ISP, and HSP blazars (LSP: $\log (\nu_{\rm peak}/\rm{Hz}) < 14$; ISP: $14 \le \log (\nu_{\rm peak}/\rm{Hz})< 15$; HSP: $\log (\nu_{\rm peak}/\rm{Hz})\ge 15$). Symbols are the same as in Figure~\ref{fig:radio_gamma_flux}.}
    \label{radio_nupeak}
\end{figure}

Figure~\ref{fig:gr_gamma_index} presents the distribution between $G_{\rm r}$ and the $\gamma$-ray photon index $\Gamma_\gamma$.
 A significant anti-correlation is observed for both BL Lacs [$\log G_{\rm{r}}=-(1.40\pm0.07)\Gamma_\gamma+(6.31\pm0.14)$ with $r=-0.55$ and $p<10^{-4}$] and FSRQs [$\log G_{\rm{r}}=-(0.67\pm0.09)\Gamma_\gamma+(4.40\pm0.21)$ with $r=-0.28$ and $p<10^{-4}$], while BL Lacs preferentially occupy regions of higher $\gamma$-ray loudness and harder spectra, and FSRQs cluster at softer photon indices and lower $G_{\rm r}$.  \citet{bock16aa} investigated the relationship between $G_{\rm r}$ and $\Gamma_\gamma$ using contemporaneous $\gamma$-ray and radio data, and reported a clear anti-correlation between the two quantities. Their result is consistent with earlier findings by \citet{linford12apj} and \citet{lister11apj}. 
The phenomenon is interpreted within the shift of the SED. As the synchrotron peak shifts to higher frequencies, the radio emission decreases, while the high-energy peak moves to higher energies, resulting in enhanced $\gamma$-ray emission and a harder spectrum \citep{bock16aa,fanxuliang18apj}. This behaviour is supported by the anti-correlation between the radio emissions and $\nu_{\rm{peak}}$, namely the blazar sequence \citep{fossati98mnras} (also shown in Fig. 
\ref{radio_nupeak}): high-$\nu_{\rm peak}$ sources are characterised by lower radio emissions. Higher radio emission (also higher radio core brightness temperature,  e.g., \citealt{linford12apj}) is generally associated with larger Doppler factors (e.g., \citealt{fan09pasj, hovatta09aa, lio18ApJ, homan21ApJ}). This implies that LBLs are more strongly beamed in the radio band than HBLs.

\citet{abdo10apj429} presented that the $\Gamma_\gamma$ is also anti-correlated with the $\nu_{\rm{peak}}$ for the 1FGL blazars.  Therefore, the relationship between $\gamma$-ray loudness and $\nu_{\rm peak}$ is expected to be positively correlated, which has been discussed in the literature \citep{ojha10A&A, lister11apj, linford12apj}. However, due to the sample limit, different studies report varying trends. The relation between the $\gamma$-ray loudness and $\nu_{\rm{peak}}$ for a large sample is presented in Sec. \ref{loud_syn}.

\subsection{$\gamma$-ray Loudness and synchrotron peak frequency}\label{loud_syn}
\begin{figure}
    \centering
    \includegraphics[width=0.95\linewidth]{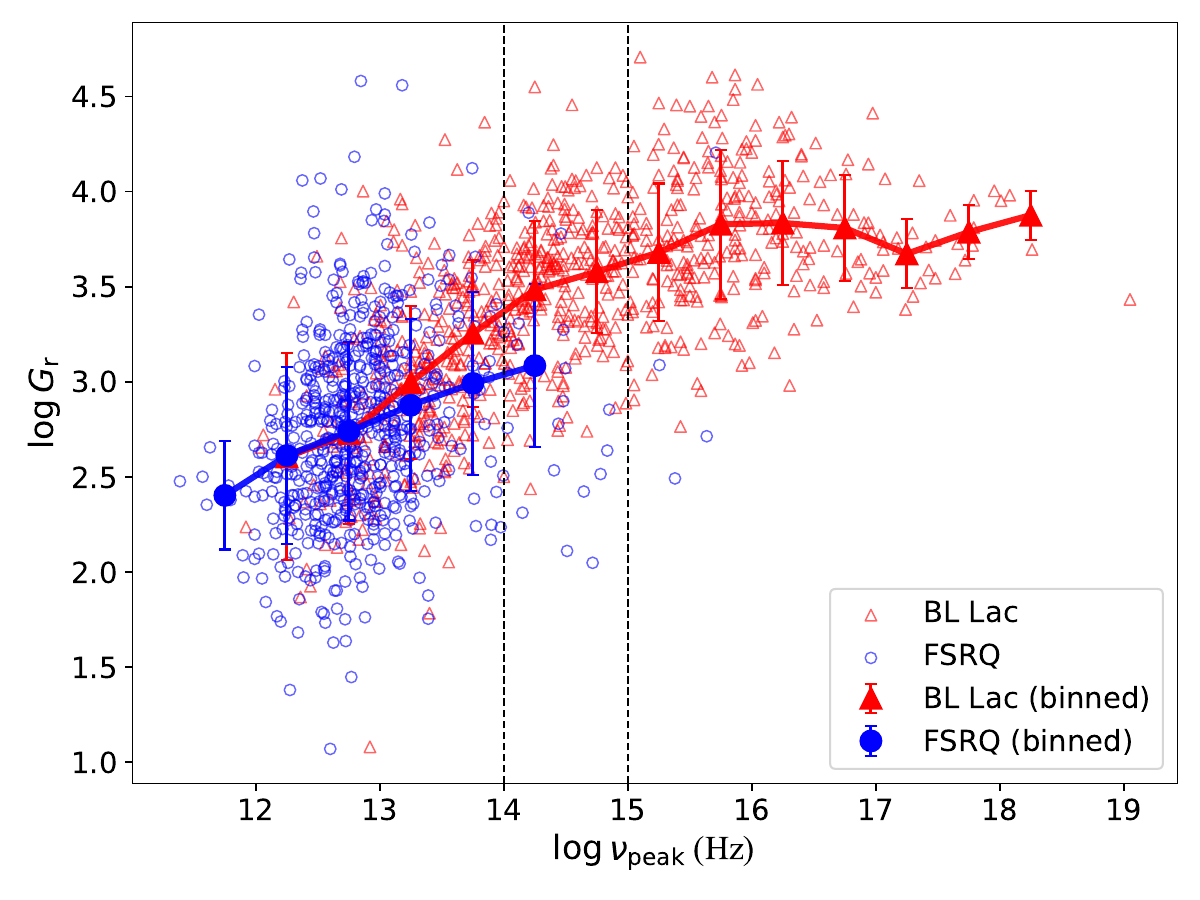}
    \caption{$\gamma$-ray loudness $G_{\rm r}$ versus synchrotron peak frequency $\nu_{\rm peak}$ for \textit{Fermi} blazars.  The interval of $\log \nu_{\rm{peak}}$ with a bin size of 0.5 for both BL Lacs and FSRQs.
    The dashed vertical lines indicate the boundaries between LSP, ISP, and HSP blazars (LSP: $\log (\nu_{\rm peak}/\rm{Hz})< 14$; ISP: $14 \le \log (\nu_{\rm peak}/\rm{Hz})< 15$; HSP: $\log (\nu_{\rm peak}/\rm{Hz})\ge 15$). Symbols are the same as in Figure~\ref{fig:radio_gamma_flux}.}
    \label{fig:gr_nupeak}
\end{figure}

\citet{ajello.2022.apjs.263} derived the $\nu_{\rm peak}$ for 1913 blazars (731 FSRQs and 1182 BL Lacs; see their Table 2) by fitting the synchrotron component of the SED with either a third-degree polynomial in the log-log plane or a parametric method based on broadband spectral indices. Among these sources, 1506 blazars (844 BL Lacs and 662 FSRQs) are matched in this paper. The distribution between $\nu_{\rm peak}$ and $G_{\rm{r}}$ for 1506 blazars is shown in Figure~\ref{fig:gr_nupeak}. The dashed vertical lines in Figure~\ref{fig:gr_nupeak} mark the boundaries between LSPs ( $\log (\nu_{\rm peak}/\rm{Hz})< 14$), ISPs ($14 \le \log (\nu_{\rm peak}/\rm{Hz})< 15$), and HSPs ($\log (\nu_{\rm peak}/\rm{Hz})\ge 15$) (e.g., \citealt{abdo10sed,fan16,yjh22apjs}). 
As expected from the relation between the $\gamma$-ray photon index and the $\nu_{\rm{peak}}$ \citep{abdo10apj429,acker11ApJ,abdollahi.2020.apjs.247}, the total tendency is positive from the lower $\nu_{\rm{peak}}$ and $\gamma$-ray loudness to the higher $\nu_{\rm{peak}}$ and $\gamma$-ray loudness. FSRQs and LBLs exhibit a weak positive trend with substantial scatter. In contrast, the trend appears to flatten in HBLs (Fig.~\ref{fig:gr_nupeak}). This phenomenon may arise from the different high-energy processes operating in these subclasses, as well as the influence of the Klein--Nishina (KN) effect.

\section{Discussion}\label{dis}
The connection between radio and $\gamma$-ray emissions in blazars has been extensively studied, particularly, since the beginning of the \textit{Fermi} era \citep{abdo10apj429,acker11ApJ741,linford11ApJ,chen11apj,lister11apj,fanxuliang2012raa,arshakian12aa,bock16aa,cui26}. The close relations between the radio and $\gamma$-ray luminosities (or fluxes) have been confirmed to support the link between the low-energy and high-energy bands. \citet{chi97aap} proposed that high-resolution VLBI data are more suitable when studying the beaming patterns through the $\gamma$-ray to radio flux ratio. \citet{fan98aa} found a close correlation between the $\gamma$-ray and the high-frequency (1.3 mm, 230 GHz) radio emissions for 44 $\gamma$-ray RLAGNs, which means that the $\gamma$-ray is associated with the radio emission from the jet.  \citet{huang99apj} investigated the correlation between the flux ratio and the Doppler factors ($\delta$) for a sample of AGNs detected by EGRET\footnote{The Energetic Gamma-Ray Experiment Telescope.}. Based on the SSC constraints proposed by \citet{ghisellini.1993.apj.407}, they estimated $\delta_{\rm{SSC}}$ and found strong correlations between $\delta_{\rm{SSC}}$ and both the $\gamma$-ray to optical flux ratio and the $\gamma$-ray to infrared flux ratio. These results favour an EC model, in which external seed photons contribute to the $\gamma$-ray emission in EGRET AGNs. However, \citet{fan17apjl} found a tight correlation between $\gamma$-ray luminosity and radio luminosities in blazars, and proposed that an SSC emission is the origin for most $\gamma$-ray blazars because EC emission is expected to dilute any tight correlation between low- and high-energy emissions, while SSC emission should show tight correlations between them.  In this paper, we compiled a large sample of \textit{Fermi} blazars, and confirmed a statistically positive correlation between the VLBI radio flux and the $\gamma$-ray energy flux (Figure~\ref{fig:radio_gamma_flux}), as also shown in \citet{abdo10apj429,pushkarev10apjl,lister11apj,arshakian12aa,linford12apj,fanxuliang2012raa,bock16aa,fan17apjl,yjh22apjs}. Such a correlation is generally interpreted as evidence that the radio and $\gamma$-ray emission originate from a one-zone emitting region within the relativistic jet. 

As pointed out by \citet{Costamante2002aap,tavecchio10mnras},
the brightest $\gamma$-ray blazars are those with the largest radio flux. Therefore, the observed correlation between the radio and $\gamma$-ray bands is likely driven by Doppler beaming, which enhances both the radio and high-energy emissions, and their correlations (e.g., \citealt{fan98aa,lister09ApJL}); however, there may be an additional dependence on the high-energy radiation mechanism. A purely EC model would not predict a tight correlation between radio and $\gamma$-ray flux if the seed photons originate entirely outside the jet. Meanwhile, the $\delta_{\rm{EC}}$ is not necessarily related to the $\delta_{\rm{syn}}$. In contrast, the SSC scenario links the synchrotron and IC emission through the same electron population and the similar inner jet Doppler factors $\delta_{\rm{SSC}}\sim\delta_{\rm{syn}}$, producing an intrinsic connection between radio and $\gamma$-ray output.   Therefore, the clear correlation between radio and $\gamma$-ray emissions favours the one-zone leptonic model with the SSC radiation mechanism and suggests that the SSC mechanism contributes to both BL Lacs and FSRQs. The SSC contributions evidently support the SEDs for some FSRQs (e.g., \citealt{ghisellini98,ghisellini2010mnras,bottcher13apj,yuan22pasp}).

Source variability and the use of non-simultaneous data can weaken the observed correlation. The averaged $\gamma$-ray flux may smooth out short-term variability, and time lags between radio and $\gamma$-ray bands may introduce additional scatter. Nevertheless, the large sample size helps reveal the underlying statistical trend. Even in FSRQs, where EC emission dominates the high-energy output, the observed correlation suggests that SSC still contributes to the radio and $\gamma$-ray components.

\subsection{SED shift for $G_{\rm{r}}$ and $\Gamma_{\gamma}$}

The $\gamma$-ray emissions for FSRQs are explained by the EC, with the higher external photon density rapidly cooling down the relativistic high-energy electrons and steepening the spectrum \citep{ghisellini17mn,elisa2022,ye26}. However, one would expect a softer $\gamma$-ray photon index and higher $\gamma$-ray emissions as presented in \citet{ghisellini11mnras} and \citet{ghisellini17mn},  where BL Lacs and FSRQs occupy different regions in the $\gamma$-ray luminosity-spectral index plane due to differences in jet power, accretion mode, and external photon fields. In this case, the stronger $\gamma$-ray emission should enhance $G_{\rm r}$ in FSRQs, and a positive correlation between $G_{\rm r}$ and $\Gamma_\gamma$ is expected from BL Lacs to FSRQs. 

\citet{lister11apj} first presented the $G_{\rm{r}}$ and $\Gamma_{\gamma}$ with an anti-correlation for both BL Lacs and FSRQs, which is confirmed and discussed by many authors (e.g., \citealt{linford12apj,fanxuliang2012raa,bock16aa,fanxuliang18apj,cui26}).  With a large sample of 1687 blazars, we also found and confirmed a significant anti-correlation between $G_{\rm{r}}$ and $\Gamma_{\gamma}$ (Figure~\ref{fig:gr_gamma_index}). As  $\nu_{\rm{peak}}$ increases, the radio flux density tends to decrease (Figure~\ref{radio_nupeak}), namely the blazar sequence \citep{fossati98mnras,elisa2022}, where weaker radio emission is observed in HSPs. Meanwhile, the shift of the high-energy SED component toward the \textit{Fermi} band enhances the observed $\gamma$-ray emission, resulting in larger $G_{\rm r}$ values. However, for sources with $\log \nu_{\rm peak} > 15$ (e.g., HSPs), the radio flux density exhibits little variation with increasing $\nu_{\rm peak}$. This behaviour may arise from the radio sensitivity limit and/or selection effects in the radio band.
Overall, it is interesting to note that the contributions from the shift of the SED might be stronger than the effect of Compton cooling for both BL Lacs and FSRQs.

\subsection{High-energy radiation mechanism for $G_{\rm{r}}$ and $\nu_{\rm{peak}}$}
   
As mentioned before, the spectral index is anti-correlated with $\nu_{\rm{peak}}$ in \citet{abdo10apj429} and \citet{acker11ApJ}. Therefore, the relationship between $G_{\rm{r}}$  and $\nu_{\rm peak}$ is expected to be closely positively correlated. But limited by the sample size, different studies report inconsistent trends:

\citet{lister11apj} proposed a parameter of the $\gamma$-ray loudness, which is the ratio of average $\gamma$-ray luminosity during the first 11 months of the \textit{Fermi} mission to the median 15 GHz VLBA radio luminosity, and showed the positive relation between the $G_{\rm{r}}$ and the $\nu_{\rm{peak}}$ for 1FGL BL Lacs. One year later, \citet{linford12apj} adopted a larger sample of the \textit{Fermi} blazars with both 1FGL blazars and 2FGL (\textit{Fermi} Second Source Catalogue) blazars, but they obtained a different result: the mostly weak or tentative correlations for BL Lacs and FSRQs were found for the 1FGL blazars, but low-significance correlations for the 2FGL sample. When a second data set from the 2LAC (the \textit{Fermi} Second Catalogue of AGN, \citealt{acker11ApJ}), estimates of the $\nu_{\rm{peak}}$ based on the empirical method (e.g., \citealt{abdo10apj429}), show even weaker correlation between the $G_{\rm{r}}$ and the $\nu_{\rm{peak}}$ \citep{linford12apj}.

As shown in Fig. 3 of \citet{linford12apj}, a handful of BL Lacs with the high $\nu_{\rm peak}$ and relatively low $G_{\rm{r}}$ significantly affected the positive correlations between the  $G_{\rm{r}}$  and $\nu_{\rm peak}$.  They explained that the different results may arise from sample selection effects. The sample used by \citet{linford12apj} contains 30 HBLs, compared to only 17 HBLs in \citet{lister11apj}. Limited by the small sample size, \citet{lister11apj} obtained a significant positive relation, which could be due to the lack of HBL samples. 

The small sample at the beginning of the \textit{Fermi} operation will preferably catch the bright sources in both the radio and $\gamma$-ray bands (e.g., \citealt{kovalev09apjl,ojha10A&A}). When a larger sample is considered in this paper, we found different trends for both FSRQs and BL Lacs, as well as the subclasses, LBLs, IBLs, and HBLs, as shown in Fig. \ref{fig:gr_nupeak}. The relation for HBLs is flatter than both LBLs and IBLs, while LBLs and IBLs do present a tentatively positive relation between the $\gamma$-ray loudness and $\nu_{\rm{peak}}$. Therefore, when more HBLs are involved in the sample, the flat trend in HBLs might naturally weaken the positive relationship.

The different slopes observed in the $G_{\rm r}$--$\nu_{\rm peak}$ relation (Figure~\ref{fig:gr_nupeak}) among LBLs, IBLs, and HBLs provide important insight into the underlying radiation processes. 
As suggested by \citet{linford12apj}, an SSC origin of the $\gamma$-ray emission would favour a strong correlation between $G_{\rm r}$ and $\nu_{\rm peak}$, since in this scenario the seed photons for IC scattering are produced internally via synchrotron emission, and both the synchrotron and $\gamma$-ray components share similar Doppler boosting factors. For FSRQs, however, the situation is more complex. The seed photon field is supplemented by external radiation components, such as those from the broad-line region or the dusty torus, leading to EC scattering. These external photon fields introduce additional Doppler boosting effects and variability in the $\gamma$-ray emission that are not directly linked to the synchrotron properties, thereby increasing the scatter in the $G_{\rm r}$--$\nu_{\rm peak}$ relation and diluting any intrinsic correlation.

In our sample of 662 FSRQs, the distribution is highly scattered, with most sources located at $\log (\nu_{\rm peak}/\rm{Hz})< 14$, consistent with an EC-dominant contribution. Nevertheless, the binned data (with bin size $\Delta \log \nu_{\rm{peak}} = 0.5$; see the binned blue circles in Fig.~\ref{fig:gr_nupeak}) reveal a weak but noticeable positive trend, in which $G_{\rm r}$ increases on average with $\nu_{\rm peak}$. This suggests that, although EC likely dominates the $\gamma$-ray emission in FSRQs, SSC may still contribute to the high-energy output and plays a role in shaping the positive correlation.

For BL Lacs, different behaviours emerge among the LBL, IBL, and HBL subclasses. Using the same binning scheme, LBLs and IBLs exhibit a clearer positive correlation between $G_{\rm r}$ and $\nu_{\rm peak}$. Meanwhile, the $G_{\rm{r}}$ of LBLs spans a wide range, from $\log G_{\rm r}=1.08$ to $\log G_{\rm r}=4.36$, making them as scattered as FSRQs in the $G_{\rm r}$--$\nu_{\rm peak}$ plane. This behaviour suggests that LBLs may resemble FSRQs, in which EC emission contributes significantly to the observed scatter, although SSC emission still plays an important role in producing the overall positive trend between $G_{\rm r}$ and $\nu_{\rm peak}$.  In contrast, IBLs exhibit a relatively tighter distribution, possibly indicating a stronger contribution from SSC emission. The enlarged samples of LBLs and IBLs used in this work improve the statistical robustness compared to earlier studies (e.g., \citealt{lister11apj,linford12apj}) and further support the importance of SSC emission in shaping the $\gamma$-ray loudness sequence.

The continuity of the $G_{\rm{r}}-\nu_{\rm{}peak}$ correlation for BL Lacs in \citet{lister11apj} suggests that the LBLs and HBLs belong to the same parent population. However, HBLs exhibit a much flatter distribution than LBLs in the $G_{\rm r}$--$\nu_{\rm peak}$ plane, with no clear correlation in this sample. A similar lack of correlation was reported for the 2FGL sample by \citet{linford12apj}, promoting that  HBLs are different from LBLs in the $G_{\rm{r}}-\nu_{\rm{}peak}$ plane. In this sample, 285 HBLs dilute the effects of the sample bias, and we proposed an explanation of the flattened trend in HBLs with the KN effect: as the characteristic electron and photon energies increase, the IC scattering becomes less efficient, suppressing the $\gamma$-ray emission and leading to the observed flattening of the relation.

\subsection{Implications for Klein-Nishina Effects}
In the standard one-zone leptonic framework for interpreting the broadband SEDs of HBLs,  when IC scattering occurs in the Thomson regime, the SSC peak frequency follows \citep{tavecchio98,chen2018apjs,fan23apjs,xiao25apj},
\begin{equation}
\nu_{\rm SSC} = \frac{4}{3} \gamma_{\rm p}^{2} \nu_{\rm syn},
\label{v_eq}
\end{equation}
where $\gamma_{\rm p}$ is the Lorentz factor of electrons that dominate the synchrotron emission. This relation implies a linear dependence between $\log \nu_{\rm SSC}$ and $\log \nu_{\rm syn}$, i.e., $\log \nu_{\rm SSC} = \log \nu_{\rm syn} + \mathrm{const.}$, with the constant determined by $\gamma_{\rm p}$. However, \citet{xiao25apj} found that the correlation between $\log \nu_{\rm SSC}$ and $\log \nu_{\rm syn}$ has a slope of $\sim 0.64$ for HBLs, deviating from the Thomson expectation. This deviation is explained by the KN effect, which becomes important when the photon energy in the electron rest frame approaches the electron rest-mass energy. In this regime, the scattering cross-section is suppressed, reducing the efficiency of the IC process and weakening the $\gamma$-ray emission.

The onset of the KN regime can be approximately described by \citep{tavecchio98},
\begin{equation}
    \gamma_{\rm b}\, \nu'_{\rm syn} \gtrsim \frac{3}{4} \frac{m_{\rm e}c^{2}}{h},
\label{KN_condi}
\end{equation}
where $m_{\rm e}$ is the electron mass, $c$ is the speed of light, and $h$ is the Planck constant, $\gamma_{\rm b}$ is the break Lorentz factor of the electron distribution, and $\nu'_{\rm syn}$ is the synchrotron peak frequency in the comoving frame, related to the observed synchrotron peak frequency $\nu_{\rm syn}$ through $\nu_{\rm syn} = \nu'_{\rm syn} \delta / (1+z)$. 
In the one-zone model, the synchrotron frequency averaged over the spectral shape for an electron of Lorentz factor $\gamma_{\rm b}$ is \citep{tavecchio98},
\begin{equation}
    \nu_{\rm syn} = 3.7 \times 10^{6} \, \gamma_{\rm b}^{2} \, B \frac{\delta}{1+z},
\label{nu_sy}
\end{equation}
where $B$ is the magnetic field strength (G) and $z$ is the redshift. Eliminating the $\gamma_{\rm b}$ by combining Eqs.~(\ref{KN_condi}) and (\ref{nu_sy}), 
\begin{equation}
    \nu_{\rm syn} \gtrsim \nu_{\rm syn}^{\rm c} = 3.17 \times 10^{15} \, B^{1/3} \frac{\delta}{1+z},
\label{nu_sy_KN_condi}
\end{equation}
which indicates that sources above this threshold are expected to experience significant KN suppression of the SSC emission.

\citet{xiao25apj} analysed the averaged broadband SEDs of \textit{Fermi} BL Lacs and derived the critical $\nu_{\rm syn}^{\rm c}$. They found that 359 out of 513 BL Lacs are affected by KN suppression. A large fraction of HBLs operate in a regime where IC scattering is significantly less efficient.

This provides a natural framework for interpreting the observed $G_{\rm r}$--$\nu_{\rm peak}$ relation. As discussed in the previous section, HBLs exhibit a much flatter distribution compared to LBLs and IBLs. Given their high $\nu_{\rm peak}$ values (note that  $\nu_{\rm syn}=\nu_{\rm{peak}}$ in this work), HBLs are more likely to satisfy the KN condition, leading to reduced IC efficiency and suppressed $\gamma$-ray emission.  As shown in Fig.~\ref{fig:gr_nupeak}, the values of $\log G_{\rm r}$ reach their plateau in a range $\log (\nu_{\rm peak}/{\rm Hz})=15.5-16$, and this behaviour may indicate the onset of KN suppression, marking the transition from the Thomson to the KN scattering regime in these sources. If this feature is indeed associated with KN effects, it can be used to constrain the magnetic field strength in the emitting region through Eq. (\ref{nu_sy_KN_condi}).

In the model involving a shock in a relativistic jet, the $\gamma$-ray emission is produced in the innermost regions of the relativistic jet, whereas the radio emission originates from the outer region \citep{2008Natur.452..966M}. Within the SSC framework, the radio and $\gamma$-ray emissions are connected through the same population of relativistic electrons, with synchrotron photons serving as seed photons for IC scattering \citep{finke2008apj}. The VLBI core emissions trace the compact, Doppler-boosted inner jet \citep{deller14,bock16aa,petrov25apjs}. Therefore, the ratio between the $\gamma$-ray and VLBI radio emissions provides a valuable probe of the relativistic beaming effect and the origin of high-energy radiation. Moreover, the radio Doppler factors reported by \citet{lio18ApJ} cover a range comparable to the $\gamma$-ray Doppler factors derived by \citet{zhang02pasj,zhang20apj,chen24apjs}, further supporting that the compact radio emission is also closely related to the strong beaming effect.
 Since $\delta$ and $B$ are inversely proportional and $\delta$ can vary widely (e.g., \citealt{hovatta09aa,lio18ApJ,chen2018apjs,ye21pasj}), $\delta$ could introduce substantial uncertainty in the estimation of the magnetic field. \citet{lian26apj} obtained the Doppler factors for 25 EHSPs in a range of $\delta=5$ and $\delta=30$ with an average of $\langle\delta\rangle=14.6$. The average Doppler factor for \textit{Fermi} BL Lacs is about $\langle\delta\rangle=15$ in \citealt{zhang20apj} and \citet{chen24apjs}. \citet{tavecchio10mnras} obtained $\delta$ in a range of $20$ to $30$ for \textit{Fermi} BL Lacs from the one-zone SSC model, but some extreme BL Lacs require $\delta>40$.  \citet{tavecchio98} investigated the magnetic field strength and Doppler factors of three TeV BL~Lacs within a homogeneous SSC framework, taking the KN effect into account. For Mrk 421, they constrained the Doppler factor from the $\gamma$--$\gamma$ pair-production opacity condition ($\delta > 15$; their Eq.~[23]), while a more stringent KN condition gives an upper limit of $\delta_{\rm KN} < 31$ (their Eq.~[17]). 

\begin{figure}
     \centering
     \includegraphics[width=1.0\linewidth]{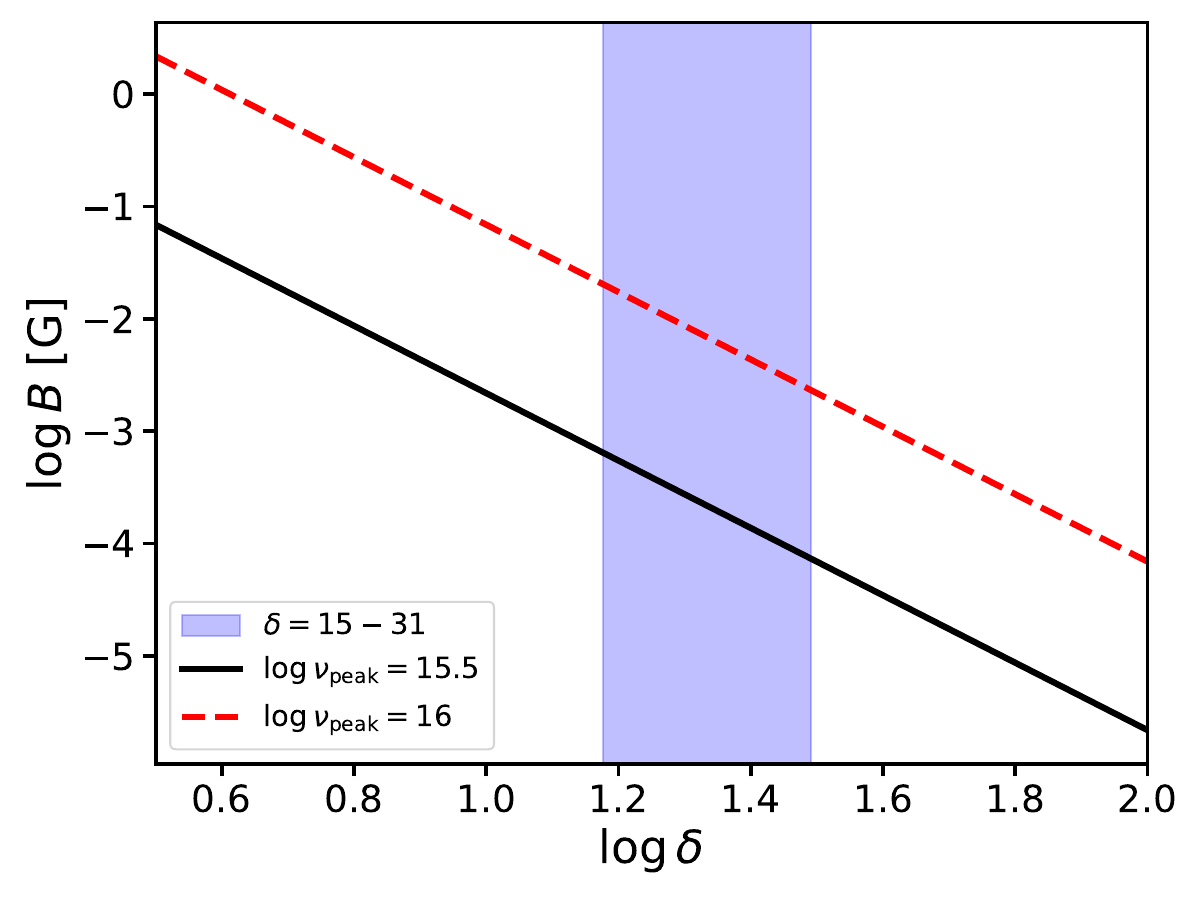}
     \caption{Magnetic field strength (B) versus Doppler factor ($\delta$). The blue shaded region indicates the range $15 < \delta < 31$. Black solid and red dashed lines denote $\log \nu_{\rm peak}=15.5$ and 16, respectively.}
     \label{delta}
 \end{figure}

Motivated by these constraints, we explored the relation between the magnetic field strength $B$ and the Doppler factor $\delta$ for HBLs with synchrotron peak frequencies in the range $\log \nu_{\rm peak}=15.5$--$16$, assuming a redshift of $z=0.3$ for BL Lacs \citep{tavecchio10mnras,ajello2020apj,ajello.2022.apjs.263}. The resulting $B$--$\delta$ relation is shown in Fig.~\ref{delta}, where the blue shaded region corresponds to the range $15 < \delta < 31$.  Although the exact onset of the KN suppression is difficult to determine observationally, a tentative plateau of $G_{\rm r}$ around $\log \nu_{\rm peak}=15.5-16$ can be seen in the HBL sample. Under this scenario, for sources ($z=0.3$) with $\log \nu_{\rm peak}$ between 15.5 and 16, the inferred magnetic field strength spans from $\log B=-4.14$ G to $\log B=-1.69$ G.

A small population of four HSP FSRQs was identified by \citet{ajello2020apj}. These sources exhibit hard $\gamma$-ray spectra similar to those of HBLs, but are located at higher redshifts. In general, \textit{Fermi} FSRQs span a much broader redshift range than BL Lacs \citep{ajello2020apj,ajello.2022.apjs.263}. According to Eq. (\ref{delta}), the inferred magnetic field strength scales as $B\propto(1+z)^3$, making it highly sensitive to redshift. For the same Doppler factor, changing the redshift from $z=0.3$ to $z=1.8$ increases the inferred magnetic field by one order of magnitude. Therefore, high-redshift HSP FSRQs entering the KN regime are expected to possess stronger magnetic fields than BL Lacs. This is consistent with the SED fitting that higher magnetic fields are more common in FSRQs than BL Lacs \citep{ghisellini98,ghisellini2010mnras,chen2018apjs,fan23apjs}.

\citet{tavecchio10mnras} analysed the BL Lacs in the \textit{Fermi} sample, including sources also detected at very high energies by Cherenkov telescopes. They found that high-energy BL Lacs can be broadly divided into two groups. The majority are well described by a standard one-zone SSC model with moderate magnetic fields and electron energies, whereas EHBLs require unusually low magnetic fields and extremely energetic electrons, highlighting the diversity of physical conditions of the emission region in BL Lacs.  \citet{fan23apjs} derived magnetic fields for 1141 BL Lacs within a one-zone leptonic framework, finding an average value of $\log B = -0.51 \pm 1.66$. Similarly, \citet{xiao25apj} obtained an average value of $\log B = 0.23 \pm 0.33$ for \textit{Fermi} BL Lacs based on physically constrained broadband SED modelling. \citet{zhao24apj} also studied the SEDs with the \textit{JetSet} for a sample of 348 \textit{Fermi} HBLs and computed that HBLs have the magnetic field in a range of $\log B=-3.66$ G to $\log B=0.60$ G with an average of $\langle\log B\rangle=-1.25$ G.
\citet{maria25aa} identified 66 new EHSP candidates and calculated the magnetic field strength based on a one-zone synchrotron/SSC framework. They proposed that  
$\nu_{\rm peak}$ reflects a more complex combination of parameters (including $\gamma_{\rm{b}}$, $B$, and $\delta$) and may capture broader physical conditions in the jet. However, they showed that sources with the lower $\nu_{\rm peak}$ have the lower magnetic field (Fig. 6 of \citealt{maria25aa}).  As discussed in \citet{maria25aa}, when the synchrotron peak shifts to higher frequencies, the energy stored in the magnetic field becomes comparable to that of the relativistic electrons, supporting a more balanced and energetically efficient jet environment in most EHBLs. \citet{lian26apj} computed the magnetic field strength ranging from $\log B=-2$ G to $\log B=-0.22$ G with an average of $\langle\log B\rangle=-1.15$ G for 25 EHSPs.  

Our estimated magnetic field strengths are broadly consistent with previous studies. The KN effect plays an important role in shaping the $\gamma$-ray loudness, particularly for HBLs. This effect can explain the flattening of the $G_{\rm r}$--$\nu_{\rm peak}$ relation and the absence of a significant correlation in the 2FGL sample reported by \citet{linford12apj}, where a larger number of HBLs were included in the analysis of the $G_{\rm{r}}$ and $\nu_{\rm{peak}}$.

\section{Conclusions}

In this work, we investigated the relation between radio and $\gamma$-ray properties for a large sample of 1687 \textit{Fermi} blazars, including an investigation of the relation between the $\nu_{\rm{peak}}$ and the $G_{\rm r}$, as well as its implications for high-energy processes. The main conclusions are as follows:

\begin{enumerate}

\item Based on a large sample of 1687 \textit{Fermi} blazars, we presented a clear positive correlation between radio and $\gamma$-ray fluxes for both BL Lacs and FSRQs. This result suggests that SSC emission contributes to the high-energy emission in both subclasses, providing a natural link between the radio and $\gamma$-ray bands.

\item Compton cooling would be expected to produce a positive correlation between $G_{\rm r}$ and $\Gamma_{\gamma}$. However, an anti-correlation is observed for the sample of 1687 \textit{Fermi} blazars. This behaviour is instead interpreted as a consequence of the SED shift, rather than being driven by Compton cooling. Therefore, the shift of the SED appears to play a dominant role in shaping the $G_{\rm r}$--$\Gamma_{\gamma}$ relation.

\item The relation between $G_{\rm r}$ and $\nu_{\rm peak}$ exhibits distinct behaviours among different blazar subclasses. FSRQs and LBLs show a weak positive trend, contributed from SSC emissions, while the substantial scatter is likely associated with EC emissions. In contrast, HBLs display a much flatter distribution. We interpreted this flattening as a consequence of the reduced IC efficiency in the KN regime. Assuming $\delta \in (15,31)$ in the SSC leptonic framework, the onset of KN suppression around $\log (\nu_{\rm peak}/{\rm Hz}) \sim 15.5$--$16$ allows one to constrain the magnetic field strength of $-4.14 < \log (B/{\rm G}) < -1.69$ for those HBLs transitioning from the Thomson to the KN regime.
\end{enumerate}

\section*{Acknowledgements}
We greatly appreciate the referee’s comments, which have helped us improve the manuscript.
We would also like to appreciate the suggestions and ideas from Dr H. B. Xiao. This work is partially supported by the National Natural Science Foundation of China (grant No. 12433004), National Key Research and Development Program of China (Grant No. 2025YFA1614102), the National SKA Program of China (No. 2025SKA0130100), the Innovation Research Team of Guangzhou University (grant No. 2023ZDP001), the Eighteenth Regular Meeting Exchange Project of The Scientific and Technological Cooperation Committee between the People’s Republic of China and the Republic of Bulgaria (Series NO. 1802). Z.Y. Pei acknowledges support from the National Science Foundation for Young Scientists of China (grant No. 12103012). We also acknowledge the science research grants from the China Manned Space Project with NO. CMS-CSST-2025-A07, and the support for Astrophysics Key Subjects of Guangdong Province. This research has made use of the NASA/IPAC Extragalactic Database (NED), which is operated by the Jet Propulsion Laboratory, California Institute of Technology, under contract with the National Aeronautics and Space Administration. This work was also supported by the Guangzhou University Graduate Student Innovation Capacity Cultivation Project under Grant No. JCCX2025010, and by the LEPL Shota Rustaveli National Science Foundation of Georgia under Grant No. FR-25-21041.  G. H. Chen (No. 202509940003) and W. X. Yang (No. 202309940006) gratefully acknowledge financial support from the China Scholarship Council.

\section*{Data Availability}
The data underlying this article are publicly available. The radio data were obtained from the RFC VLBI catalogue (DOI: 10.25966/dhrk-zh08), and the $\gamma$-ray data were obtained from the \textit{Fermi} 4LAC-DR3 AGN catalogue (DOI: https://doi.org/10.26093/cds/vizier.22630024). Additional data products generated during this study are available from the corresponding author upon reasonable request.  
The complete Table 1 underlying this article is available in Zenodo, at https://doi.org/10.5281/zenodo.21695209 [doi].
%%%%%%%%%%%%%%%%%%%%%%%%%%%%%%%%%%%%%%%%%%%%%%%%%%



\bibliographystyle{mnras}
\bibliography{ye_bib}
% if your bibtex file is called example.bib

%\begin{thebibliography}{99}
%Author A.\simN., 2013, Journal of Improbable Astronomy, 1, 1
%\end{thebibliography}

%%%%%%%%%%%%%%%%%%%%%%%%%%%%%%%%%%%%%%%%%%%%%%%%%%

%%%%%%%%%%%%%%%%% APPENDICES %%%%%%%%%%%%%%%%%%%%%

%\appendix

%The whole Table 1:
%If you want to present additional material that would interrupt the flow of the main paper,

%%%%%%%%%%%%%%%%%%%%%%%%%%%%%%%%%%%%%%%%%%%%%%%%%%

% Don't change these lines
\bsp	% typesetting comment
\label{lastpage}
\end{document}